\documentclass[twocolumn,letterpaper,aps,prl,superscriptaddress,showpacs,floatfix]{revtex4}
\usepackage{graphicx}	% Include figure filed

\usepackage{xspace}	% Include xspace

\newcommand{\pt}{\mbox{$p_T$}\xspace}
\newcommand{\mt}{\mbox{$m_T$}\xspace}

\newcommand{\raa}{\mbox{$R_{\rm AA}$}\xspace}
\newcommand{\rda}{\mbox{$R_{d{\rm A}}$}\xspace}

\newcommand{\Ncoll}{\mbox{$N_{\rm coll}$}\xspace}

\newcommand{\sqsn}{\mbox{$\sqrt{s_{_{NN}}}$}\xspace}

\begin{document}


\title{Cold-nuclear-matter effects on heavy-quark production in $d$+Au 
collisions at $\sqrt{s_{_{NN}}}$=200 GeV}

\newcommand{\abilene}{Abilene Christian University, Abilene, Texas 79699, USA}
\newcommand{\banaras}{Department of Physics, Banaras Hindu University, Varanasi 221005, India}
\newcommand{\barc}{Bhabha Atomic Research Centre, Bombay 400 085, India}
\newcommand{\bnlcoll}{Collider-Accelerator Department, Brookhaven National Laboratory, Upton, New York 11973-5000, USA}
\newcommand{\bnlphys}{Physics Department, Brookhaven National Laboratory, Upton, New York 11973-5000, USA}
\newcommand{\caucr}{University of California - Riverside, Riverside, California 92521, USA}
\newcommand{\charlesczech}{Charles University, Ovocn\'{y} trh 5, Praha 1, 116 36, Prague, Czech Republic}
\newcommand{\chonbuk}{Chonbuk National University, Jeonju, 561-756, Korea}
\newcommand{\ciae}{Science and Technology on Nuclear Data Laboratory, China Institute of Atomic Energy, Beijing 102413, P.~R.~China}
\newcommand{\cns}{Center for Nuclear Study, Graduate School of Science, University of Tokyo, 7-3-1 Hongo, Bunkyo, Tokyo 113-0033, Japan}
\newcommand{\colorado}{University of Colorado, Boulder, Colorado 80309, USA}
\newcommand{\columbia}{Columbia University, New York, New York 10027 and Nevis Laboratories, Irvington, New York 10533, USA}
\newcommand{\czechtech}{Czech Technical University, Zikova 4, 166 36 Prague 6, Czech Republic}
\newcommand{\dapnia}{Dapnia, CEA Saclay, F-91191, Gif-sur-Yvette, France}
\newcommand{\elte}{ELTE, E{\"o}tv{\"o}s Lor{\'a}nd University, H - 1117 Budapest, P{\'a}zm{\'a}ny P. s. 1/A, Hungary}
\newcommand{\ewha}{Ewha Womans University, Seoul 120-750, Korea}
\newcommand{\fit}{Florida Institute of Technology, Melbourne, Florida 32901, USA}
\newcommand{\fsu}{Florida State University, Tallahassee, Florida 32306, USA}
\newcommand{\gsu}{Georgia State University, Atlanta, Georgia 30303, USA}
\newcommand{\hiroshima}{Hiroshima University, Kagamiyama, Higashi-Hiroshima 739-8526, Japan}
\newcommand{\ihepprot}{IHEP Protvino, State Research Center of Russian Federation, Institute for High Energy Physics, Protvino, 142281, Russia}
\newcommand{\illuiuc}{University of Illinois at Urbana-Champaign, Urbana, Illinois 61801, USA}
\newcommand{\inrras}{Institute for Nuclear Research of the Russian Academy of Sciences, prospekt 60-letiya Oktyabrya 7a, Moscow 117312, Russia}
\newcommand{\instpasczech}{Institute of Physics, Academy of Sciences of the Czech Republic, Na Slovance 2, 182 21 Prague 8, Czech Republic}
\newcommand{\isu}{Iowa State University, Ames, Iowa 50011, USA}
\newcommand{\jyvaskyla}{Helsinki Institute of Physics and University of Jyv{\"a}skyl{\"a}, P.O.Box 35, FI-40014 Jyv{\"a}skyl{\"a}, Finland}
\newcommand{\kek}{KEK, High Energy Accelerator Research Organization, Tsukuba, Ibaraki 305-0801, Japan}
\newcommand{\korea}{Korea University, Seoul, 136-701, Korea}
\newcommand{\kurchatov}{Russian Research Center ``Kurchatov Institute", Moscow, 123098 Russia}
\newcommand{\kyoto}{Kyoto University, Kyoto 606-8502, Japan}
\newcommand{\labllr}{Laboratoire Leprince-Ringuet, Ecole Polytechnique, CNRS-IN2P3, Route de Saclay, F-91128, Palaiseau, France}
\newcommand{\lawllnl}{Lawrence Livermore National Laboratory, Livermore, California 94550, USA}
\newcommand{\losalamos}{Los Alamos National Laboratory, Los Alamos, New Mexico 87545, USA}
\newcommand{\lpc}{LPC, Universit{\'e} Blaise Pascal, CNRS-IN2P3, Clermont-Fd, 63177 Aubiere Cedex, France}
\newcommand{\lund}{Department of Physics, Lund University, Box 118, SE-221 00 Lund, Sweden}
\newcommand{\maryland}{University of Maryland, College Park, Maryland 20742, USA}
\newcommand{\mass}{Department of Physics, University of Massachusetts, Amherst, Massachusetts 01003-9337, USA }
\newcommand{\muenster}{Institut fur Kernphysik, University of Muenster, D-48149 Muenster, Germany}
\newcommand{\muhlenberg}{Muhlenberg College, Allentown, Pennsylvania 18104-5586, USA}
\newcommand{\myongji}{Myongji University, Yongin, Kyonggido 449-728, Korea}
\newcommand{\nagasaki}{Nagasaki Institute of Applied Science, Nagasaki-shi, Nagasaki 851-0193, Japan}
\newcommand{\newmex}{University of New Mexico, Albuquerque, New Mexico 87131, USA }
\newcommand{\nmsu}{New Mexico State University, Las Cruces, New Mexico 88003, USA}
\newcommand{\ornl}{Oak Ridge National Laboratory, Oak Ridge, Tennessee 37831, USA}
\newcommand{\orsay}{IPN-Orsay, Universite Paris Sud, CNRS-IN2P3, BP1, F-91406, Orsay, France}
\newcommand{\peking}{Peking University, Beijing 100871, P.~R.~China}
\newcommand{\pnpi}{PNPI, Petersburg Nuclear Physics Institute, Gatchina, Leningrad region, 188300, Russia}
\newcommand{\riken}{RIKEN Nishina Center for Accelerator-Based Science, Wako, Saitama 351-0198, Japan}
\newcommand{\rikjrbrc}{RIKEN BNL Research Center, Brookhaven National Laboratory, Upton, New York 11973-5000, USA}
\newcommand{\rikkyo}{Physics Department, Rikkyo University, 3-34-1 Nishi-Ikebukuro, Toshima, Tokyo 171-8501, Japan}
\newcommand{\saispbstu}{Saint Petersburg State Polytechnic University, St. Petersburg, 195251 Russia}
\newcommand{\saopaulo}{Universidade de S{\~a}o Paulo, Instituto de F\'{\i}sica, Caixa Postal 66318, S{\~a}o Paulo CEP05315-970, Brazil}
\newcommand{\stonybrkc}{Chemistry Department, Stony Brook University, SUNY, Stony Brook, New York 11794-3400, USA}
\newcommand{\stonycrkp}{Department of Physics and Astronomy, Stony Brook University, SUNY, Stony Brook, New York 11794-3400, USA}
\newcommand{\tenn}{University of Tennessee, Knoxville, Tennessee 37996, USA}
\newcommand{\titech}{Department of Physics, Tokyo Institute of Technology, Oh-okayama, Meguro, Tokyo 152-8551, Japan}
\newcommand{\tsukuba}{Institute of Physics, University of Tsukuba, Tsukuba, Ibaraki 305, Japan}
\newcommand{\vandy}{Vanderbilt University, Nashville, Tennessee 37235, USA}
\newcommand{\waseda}{Waseda University, Advanced Research Institute for Science and Engineering, 17 Kikui-cho, Shinjuku-ku, Tokyo 162-0044, Japan}
\newcommand{\weizmann}{Weizmann Institute, Rehovot 76100, Israel}
\newcommand{\wigner}{Institute for Particle and Nuclear Physics, Wigner Research Centre for Physics, Hungarian Academy of Sciences (Wigner RCP, RMKI) H-1525 Budapest 114, POBox 49, Budapest, Hungary}
\newcommand{\yonsei}{Yonsei University, IPAP, Seoul 120-749, Korea}
\affiliation{\abilene}
\affiliation{\banaras}
\affiliation{\barc}
\affiliation{\bnlcoll}
\affiliation{\bnlphys}
\affiliation{\caucr}
\affiliation{\charlesczech}
\affiliation{\chonbuk}
\affiliation{\ciae}
\affiliation{\cns}
\affiliation{\colorado}
\affiliation{\columbia}
\affiliation{\czechtech}
\affiliation{\dapnia}
\affiliation{\elte}
\affiliation{\ewha}
\affiliation{\fit}
\affiliation{\fsu}
\affiliation{\gsu}
\affiliation{\hiroshima}
\affiliation{\ihepprot}
\affiliation{\illuiuc}
\affiliation{\inrras}
\affiliation{\instpasczech}
\affiliation{\isu}
\affiliation{\jyvaskyla}
\affiliation{\kek}
\affiliation{\korea}
\affiliation{\kurchatov}
\affiliation{\kyoto}
\affiliation{\labllr}
\affiliation{\lawllnl}
\affiliation{\losalamos}
\affiliation{\lpc}
\affiliation{\lund}
\affiliation{\maryland}
\affiliation{\mass}
\affiliation{\muenster}
\affiliation{\muhlenberg}
\affiliation{\myongji}
\affiliation{\nagasaki}
\affiliation{\newmex}
\affiliation{\nmsu}
\affiliation{\ornl}
\affiliation{\orsay}
\affiliation{\peking}
\affiliation{\pnpi}
\affiliation{\riken}
\affiliation{\rikjrbrc}
\affiliation{\rikkyo}
\affiliation{\saispbstu}
\affiliation{\saopaulo}
\affiliation{\stonybrkc}
\affiliation{\stonycrkp}
\affiliation{\tenn}
\affiliation{\titech}
\affiliation{\tsukuba}
\affiliation{\vandy}
\affiliation{\waseda}
\affiliation{\weizmann}
\affiliation{\wigner}
\affiliation{\yonsei}
\author{A.~Adare} \affiliation{\colorado}
\author{C.~Aidala} \affiliation{\mass}
\author{N.N.~Ajitanand} \affiliation{\stonybrkc}
\author{Y.~Akiba} \affiliation{\riken} \affiliation{\rikjrbrc}
\author{H.~Al-Bataineh} \affiliation{\nmsu}
\author{J.~Alexander} \affiliation{\stonybrkc}
\author{A.~Angerami} \affiliation{\columbia}
\author{K.~Aoki} \affiliation{\kyoto} \affiliation{\riken}
\author{N.~Apadula} \affiliation{\stonycrkp}
\author{Y.~Aramaki} \affiliation{\cns} \affiliation{\riken}
% ...
\collaboration{PHENIX Collaboration} \noaffiliation

\date{\today}

%------------------------------------------------------------------------------|

\begin{abstract}

The PHENIX experiment has measured electrons and positrons at midrapidity 
from the decays of hadrons containing charm and bottom quarks produced in 
$d+$Au and $p$$+$$p$ collisions at $\sqrt{s_{_{NN}}}$ = 200~GeV in the 
transverse-momentum range $0.85{\leq}p_T^{e}{\leq}8.5$ GeV/$c$.  In 
central $d$+Au collisions, the nuclear modification factor $R_{dA}$ at 
$1.5<{p_T}<5$ GeV/$c$ displays evidence of enhancement of these electrons, 
relative to those produced in $p$$+$$p$ collisions, and shows that the 
mass-dependent Cronin enhancement observed at the Relativistic Heavy Ion 
Collider extends to the heavy $D$ meson family.  A comparison with the 
neutral-pion data suggests that the difference in cold-nuclear-matter 
effects on light- and heavy-flavor mesons could contribute to the observed 
differences between the $\pi^{0}$ and heavy-flavor-electron nuclear 
modification factor $R_{\rm AA}$.

\end{abstract}

	
\maketitle

%\textbf{*** page break for PRL word count ***}  \clearpage

The experimental collaborations at the Relativistic Heavy Ion Collider 
(RHIC) have established that a hot, dense medium with partonic degrees of 
freedom is formed in Au$+$Au collisions at \sqsn = 
200~GeV~\cite{PHENIXwhite,white2,white3,white4}.  The temperature 
achieved in this medium, as inferred from direct-photon measurements, is 
well over the threshold expected from lattice-quantum-chromodynamics (QCD) 
calculations to enable deconfinement and create the quark gluon plasma 
\cite{PPG086}.  Studies of the interactions of heavy quarks with this 
matter are of particular interest.  Since charm and bottom quarks are 
dominantly produced by gluon fusion in the early stages of the collision, 
they experience the complete evolution of the system.  The 
heavy-quark-production baseline in $p$$+$$p$ collisions at \sqsn = 200 GeV 
is consistent with fixed order plus next-to-leading-log perturbative QCD 
calculations within uncertainties \cite{PPG065}.  In central Au$+$Au 
collisions, suppression of electrons from the decays of hadrons containing 
heavy quarks has been measured relative to the yield in $p$$+$$p$ scaled 
by the number of nucleon-nucleon collisions, \Ncoll, suggesting that heavy 
quarks lose a significant amount of their initial energy \cite{PPG066}.  
The positive elliptic flow amplitude of these decay electrons implies that 
heavy quarks flow along with the light partons that compose the bulk of 
the medium.  When considered together, the suppression and elliptic flow 
of these quarks are in qualitative agreement with calculations based on 
Langevin transport models that imply a viscosity to entropy density ratio 
close to the conjectured quantum lower bound of 1/4$\pi$ \cite{Teaney, 
vanHees, etaovers}.

A full understanding of these phenomena requires measurements of cold 
nuclear matter (CNM) effects, which are believed to be present in Au$+$Au 
collisions but are difficult to distinguish experimentally from effects 
due to interactions with the hot medium.  Initial state effects, such as 
modification of the parton distribution functions in the nucleus, momentum 
broadening due to parton scattering in the nucleus, and CNM energy loss 
will affect heavy-quark production rates and cannot be accounted for with 
a reference from $p$$+$$p$ data \cite{vogt_kt,EPS09,CNM_Eloss}.  It is 
therefore necessary to study $p$+Au (or $d$+Au) collisions, where a hot 
nuclear medium is not expected to form, to isolate these nuclear effects.

To this end, a vigorous experimental effort to quantify CNM effects is 
underway at RHIC.  A mass-dependent Cronin enhancement has been observed 
for $\pi, K,$ and $p$ production \cite{PPG030,STAR_Cronin}, where the 
$p_{T}$ spectra of these hadrons in $d+$Au collisions are hardened with 
respect to $p$$+$$p$.  While overall $J/\psi$ production is suppressed in 
$d+$Au collisions, a broadening of the $\pt$ spectrum is also observed 
\cite{PPG125, PPG109}.  The relative strengths and centrality dependence 
of initial-state effects and breakup in the cold nuclear medium that 
contribute to these phenomena are not known. The study of mesons 
containing open heavy flavor can help disentangle these coexisting 
effects.  This Letter presents measurements of \pt spectra and the nuclear 
modification factor (\rda) of electrons and positrons from the decays of 
hadrons containing charm and bottom quarks ($e^{\pm}_{\rm HF}$) produced 
in $d+$Au collisions at \sqsn = 200 GeV.  When combined with heavy-quark 
measurements from $p$$+$$p$ and Au$+$Au, this analysis provides a detailed 
study of the production of heavy quarks, the effects of production in a 
nucleus, and the dynamics of the hot nuclear medium.

The PHENIX experiment \cite{PHENIXNIM} sampled 80 nb$^{-1}$ of integrated 
luminosity during the 2008 $d+$Au run at RHIC, a factor of 30 increase 
over the 2003 $d+$Au data set.  The minimum bias (MB) trigger and event 
centrality are obtained from two beam-beam counters located at $3.1 <
|\eta| < 3.9$ in pseudorapidity.  The charge generated in the 
beam-beam counter facing the incoming Au nucleus is divided into four 
categories covering the 0--20\%, 20--40\%, 40--60\%, and 60--88\% most 
central collisions.  As the MB-trigger efficiency is 88$\pm$4\% of the 
total $d+$Au inelastic cross section, a correction factor is applied to 
the yield measured in the MB-triggered data sample to give a nonbiased 
sample, covering 100\% of the $d+$Au collision centrality.

This analysis considers electrons and positrons identified in the two 
PHENIX central arm spectrometers.  Each arm covers an azimuthal angle 
$\Delta\phi = \pi/2$ and a pseudorapidity range $|\eta| < 0.35$, and uses 
layers of multiwire proportional chambers and pad chambers for charged 
particle tracking.  Ring-imaging \v{C}erenkov (RICH) counters and 
electromagnetic calorimeters (EMCal) provide electron-identification and 
hadron-rejection capabilities.  A coincidence of the MB trigger and a RICH 
hit matched with an energy deposit of at least 600 or 800 MeV in the EMCal 
functions as an electron trigger.  At $\pt = 5$ GeV/$c$, charged pions 
begin to radiate in the RICH, but matching requirements between the 
track's energy deposit in the EMCal and reconstructed momentum effectively 
eliminate hadron contamination out to $\pt = 8$ GeV/$c$.  Above this, 
hadronic contamination accounts for $20\pm10\%$ of the signal, and is 
subtracted.  A full {\sc geant} simulation of the PHENIX detector is used 
to correct for the incomplete azimuthal acceptance and 
electron-identification efficiency of the central-arm detectors.

Most of the electrons produced in collisions at RHIC come not from 
heavy-flavor decays, but from the neutral-pion Dalitz decay, 
$\pi^{0}\rightarrow\gamma e^{+}e^{-}$.  The $\eta$ Dalitz decay 
contributes about 10$\%$ of the electron background for $1 < \pt < 9 $ 
GeV/$c$.  Other hadron decays ($\eta', \rho,\omega,\phi, \Upsilon$) add to 
the background at the few percent level.  Internal and external 
conversions of direct photons, while negligible at $\pt < 2 $ GeV/$c$, are 
significant sources of electrons at high momentum.  Electrons from the 
decay $J/\psi \rightarrow e^{+}e^{-}$ are a significant source of 
background at intermediate $\pt$, and constitute a maximum of about 25$\%$ 
of the total electron background at $\pt = 5$ GeV/$c$.  Conversions of 
photons from hadron decays are significant at all momenta, however, the 
low material design of the PHENIX detector ensures that the number of 
these conversion electrons is less than half of that from neutral-pion 
Dalitz decay.  In addition, electrons produced at displaced vertices from 
the $K_{e3}$ decays of $K$ mesons are misreconstructed by the PHENIX 
tracking algorithm and contribute about $3\%$ of the total background at 
$\pt = 0.85$ GeV/$c$, but quickly fall off to less than 1$\%$ at $\pt = 
1.5$ GeV/$c$.

Two independent methods are used to isolate the contribution of heavy 
flavor electrons. The cocktail method uses a Monte Carlo hadron decay 
generator to calculate the electron background from each relevant hadron 
species.  The parametrization of the neutral-pion $\pt$ spectrum is 
determined by a modified Hagedorn fit to pion data obtained from earlier 
measurements in $d+$Au \cite{PPG030,PPG044}.  The shape of the $\pt$ 
spectra of the other mesons is determined by $\mt$ scaling the pion fit, 
that is, the variable substitution $\pt \rightarrow \mt = \sqrt{\pt^{2} + 
(M_{\rm meson}^{2}-m_{\pi^{0}}^{2})}$, and their normalization is set to 
world averages of the ratio of meson/$\pi^{0}$ at high momentum 
\cite{PPG044,PDG}.  Direct-photon contributions are estimated by scaling 
the measured direct-photon yield in $p$$+$$p$ by \Ncoll~\cite{PPG060}.  
The number of conversion electrons is found by a full {\sc geant} 
simulation of the PHENIX detector material, and a similar simulation, in 
conjunction with the actual PHENIX tracking algorithm, is used to estimate 
the $K_{e3}$ decay background.  Contributions from $J/\psi$ decays are 
found by parameterizing the measured $J/\psi$ spectrum from \cite{PPG125} 
for each centrality, for $d+$Au, and from \cite{PPG097} for $p$$+$$p$.  
The small background due to $\Upsilon$ decays and the Drell-Yan process 
are taken from \cite{PPG077}, and scaled by \Ncoll for each centrality.  
The sum of these background sources is then subtracted from the inclusive 
electron measurement to give the heavy flavor contribution.

The second method of signal extraction is based on the fact that the vast 
majority of the background electrons are ``photonic" in nature, i.e. they 
originate from either a real photon (the conversion electrons) or a 
virtual photon (the electrons from Dalitz decays), while signal electrons 
are nonphotonic.  The inclusive yield of electrons in the standard 
detector configuration can be parametrized as

\begin{equation}
\textit{$N^{\rm standard}_{e} = N^{\gamma} + N^{{\rm non}\gamma}$ }
\label{eqn:conv_out}
\end{equation}

\noindent where $N^{\gamma}$ ($N^{{\rm non}\gamma}$) represents the 
photonic (nonphotonic) electron yield.  The addition of extra material 
(the ``converter", a sheet of brass 1.68$\%$ of a radiation length thick, 
wrapped around the beam pipe) into the PHENIX aperture increases the 
photonic component by a factor $R_{\gamma}$, but attenuates the signal by 
an amount (1-$\epsilon$), giving a total yield

\begin{equation}
\textit{$N^{\rm converter}_{e}= R_{\gamma}N^{\gamma} 
+ (1-\epsilon)N^{{\rm non}\gamma}$ }
\label{eqn:conv_in}
\end{equation}

By modeling the converter material in simulation, the factors $R_{\gamma}$ 
and $\epsilon$ are determined to be $2.32\pm2.7\%$ (with a slight $\pt$ 
dependence), and $0.021\pm25\%$, respectively.  The inclusive yields 
$N^{\rm standard}_{e}$ and $N^{\rm converter}_{e}$ are measured by the 
PHENIX spectrometer, so a simultaneous solution of Eqs. (1) and (2) gives 
the quantity of interest $N^{{\rm non}\gamma}$.  The nonphotonic 
background sources, namely $K_{e3}$ decays and the dielectron decays of 
the $\rho, \omega$, $\phi$, $J/\psi$, and $\Upsilon$ contribute about 
$10\%$ of the total background at $\pt < 1$ GeV/$c$, and are subtracted 
following the cocktail method described above.  The converter method 
provides a robust but statistics-limited determination of the photonic 
background.  Since the converter material creates an undesirable 
background for other measurements, only 3$\%$ of the $d+$Au data recorded 
by PHENIX in 2008 was taken with the converter installed.

\begin{figure}[t]
	\centering
\includegraphics[width=1.0\linewidth]{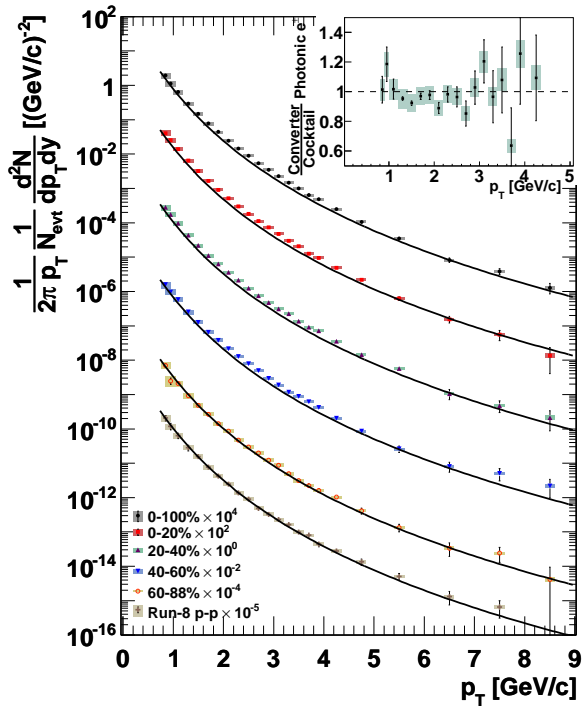}
\caption{(color online)
Electrons from heavy flavor decays, separated by centrality.  The lines 
represent a fit to the previous $p$$+$$p$ result \cite{PPG077}, scaled by 
\Ncoll.  The inset shows the ratio of photonic background electrons 
determined by the converter and cocktail methods for Minimum Bias 
$d+$Au collisions, with error bars (boxes) that represent the statistical 
uncertainty on the converter data (systematic uncertainty on the 
photonic-electron cocktail).  See text for details on uncertainties.
}
\label{fig:spectra_example}
\end{figure}

A crucial cross-check of this measurement's accuracy is the consistency of 
these two independent background determination methods.  A comparison of 
the photonic components of the cocktail (Dalitz decay electrons, 
conversions, and direct photons) to the photonic-electron signal extracted 
by the converter method shows agreement within 8\% for all centralities 
(see inset of Fig.~\ref{fig:spectra_example}).  Since the converter method 
gives a direct measurement of the photonic background, while the cocktail 
is a calculation that relies on simulation, the photonic components of the 
cocktail are scaled to match the converter data in each centrality by 
factors ranging from 0.92 to 1.01.  Detailed descriptions of these methods 
can be found in \cite{PPG077}.

%%%%%%%%%%%%%%%%%%%%%%%%%%%%%%%%%%%%%%%%%%%%%% Fig_1

%%%%%%%%%%%%%%%%%%%%%%%%%%%%%%%%%%%%%%%%%%%%%% Fig_2
\begin{figure}[t]
	\centering
\includegraphics[width=1.0\linewidth]{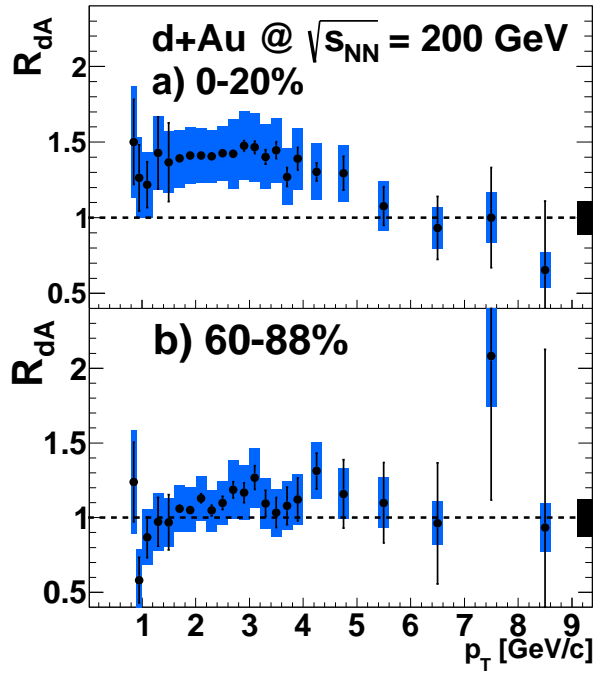}
\caption{(color online)
The nuclear modification factor, \rda, for electrons from open heavy 
flavor decays, for the (a) most central and (b) most peripheral 
centrality bins.
}
	\label{fig:RdAs}
\end{figure}

Figure~\ref{fig:spectra_example} shows the $\pt$ spectrum of electrons 
from open heavy flavor decays for each $d+$Au centrality bin, and for 
$p$$+$$p$ collisions that were measured during the same RHIC Run period 
with identical techniques.  The heavy flavor electron yield is determined 
by the cocktail method, with photonic components scaled to match the 
converter data.  The statistical (systematic) uncertainties are shown as 
bars (boxes) around the central values.  The boxes contain the 
uncertainties in the solid angle correction, electron-identification 
efficiency, and trigger-bias correction.  Added in quadrature with those 
is the uncertainty from the cocktail subtraction.  The lines are a FONLL 
spectral shape \cite{pp_fit} fitted to a previous $p$$+$$p$ heavy-flavor 
electron measurement \cite{PPG077}, scaled by \Ncoll for each 
centrality.  The $p$$+$$p$ data presented here are in good agreement with 
our previous $p$$+$$p$ results, however, the statistical uncertainties on 
the new data are $\sim2 \times$ larger.  Fitting a constant to the ratio 
of the new data to the old yields a value of 0.97 $\pm$ 0.02, with 
$\chi^{2}$/n.d.f = 20.3/26.  The fact that the 2008 $p$$+$$p$ data agree 
with the previous $p$$+$$p$ data provides an important cross-check on the 
methods used to extract the 2008 $d+$Au $e^{\pm}_{\rm HF}$ spectra.

Due to changes in the detector configuration that resulted in increased 
photon conversion background at low $\pt$, the signal to background at low 
$\pt$ is not as good as it was in previous measurements. Coupled with the 
fact that $\sim$ 90 $\%$ of the electrons from charmed hadron decays fall 
below $\pt$ = 0.8 GeV/$c$, where the present data cut off, this means that 
the data do not place meaningful constraints on the total charm production 
cross section.

%%%%%%%%%%%%%%%%%%%%%%%%%%%%%%%%%%%%%%%%%%%%%% Fig_3
\begin{figure}[t]
	\centering
\includegraphics[width=1.0\linewidth]{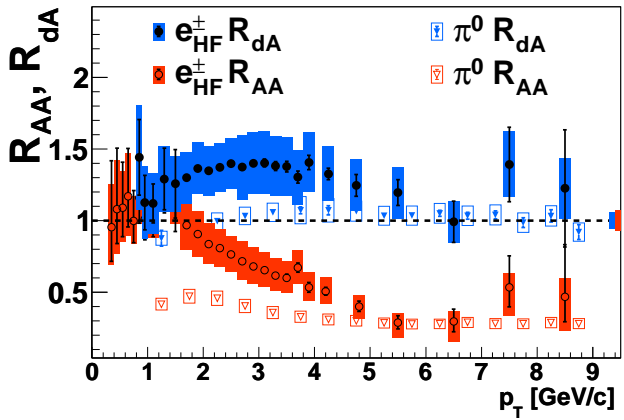}
\caption{(color online)
The nuclear modification factors \rda and \raa for minimum bias $d+$Au and 
Au$+$Au collisions, for the $\pi^{0}$ and $e^{\pm}_{\rm HF}$.  The two 
boxes on the right side of the plot represent the global uncertainties in 
the $d+$Au (left) and Au$+$Au (right) values of \Ncoll.  An additional 
common global scaling uncertainty of 9.7$\%$ on \rda and \raa from the 
$p$$+$$p$ reference data is omitted for clarity.
}
\label{fig:Rmod}
\end{figure}

The $d+$Au electron spectra are directly compared to the $p$$+$$p$ 
reference data by computing 

\begin{equation}
R_{d{\rm A}} = \frac{dN^{e}_{d{\rm A}}/dp_T}
{\langle N_{\rm coll} \rangle \times dN^{e}_{pp}/dp_T}
\label{eqn:rda}
\end{equation}
 
\noindent for each centrality.  Figure~\ref{fig:RdAs} shows \rda as a 
function of $\pt$ for the most-peripheral and most-central centrality 
bins.  As in Fig.~\ref{fig:spectra_example}, the statistical (systematic) 
uncertainties are represented by bars (boxes).  For points at $\pt < 1.6$ 
GeV/$c$, \rda is found by dividing point-by-point the $d+$Au yield by the 
$p$$+$$p$ yield from~\cite{PPG077}.  At higher transverse momentum, where 
the $p$$+$$p$ heavy-flavor electron spectrum is consistent with a shape 
from pQCD, a fit to the spectral shape from the~\cite{pp_fit} calculations 
is used to represent the $p$$+$$p$ yield.  The statistical 
uncertainty on the fit is included as a systematic uncertainty on the 
shape of \rda by adding it in quadrature with the systematic 
uncertainties on the electron background subtraction and solid angle and 
efficiency corrections.  The global scaling uncertainty from the 
uncertainty in \Ncoll and the total sampled $p$$+$$p$ luminosity is 
given by a box on the right.  Note that the 2008 $p$$+$$p$ data shown in 
Fig.~\ref{fig:spectra_example} could be used for the denominator of 
\rda, however, the use of the more precise data from \cite{PPG077} gives 
a smaller uncertainty on \rda.

The central \rda shows an enhancement out to $\pt \approx 5$ GeV/$c$, 
and implies that the suppression of heavy flavor electrons in central 
Au$+$Au collisions at RHIC is not an initial state CNM effect, but rather 
is due to the hot nuclear medium.  The peripheral nuclear modification 
factor also shows some evidence of an enhancement, which is to be expected 
since even the most peripheral centrality bin in $d+$Au samples a 
significant nuclear thickness.  Although the techniques used here do not 
allow separation of electrons from charm and bottom decays from each 
other, measurements from $p$$+$$p$ show that $\pt = 5$ GeV/$c$ is near the 
transition point where contributions from bottom quarks begin to dominate 
over charm \cite{PPG094}.  Since the total charm cross section is expected 
to scale with \Ncoll, this enhancement below 5 GeV/$c$ suggests a $\pt$ 
broadening of the $D$ spectral shape, with a mass dependence that roughly 
follows the previously observed trend in the $\pi, K,$ and $p$ families.  
The $B$ spectrum may also be modified, however, the uncertainties on the 
data and on the relative $D$ and $B$ contributions to the electron spectra 
preclude a precise determination of any effects.

The effects of cold nuclear matter are expected to be present in the 
initial state of A+A collisions, however, this CNM enhancement is 
convolved with the suppressing effects of hot nuclear matter.  
Figure~\ref{fig:Rmod} shows \rda and \raa for $e^{\pm}_{\rm HF}$ and the 
neutral pion, for which only small CNM effects are observed \cite{PPG044, 
PPG080}.  Above $\pt \approx 5$ GeV/$c$, where the CNM effects on both 
species are small, their \raa values are consistent within uncertainties.  
However, in the range where CNM enhancement is large for $e^{\pm}_{\rm 
HF}$ and small on $\pi^{0}$, the corresponding $e^{\pm}_{\rm HF} \raa$ 
values are consistently above the $\pi^{0}$ values.  This could suggest 
that the difference in the initial state cold nuclear matter effects due 
to the mass-dependent Cronin enhancement is reflected in the final state 
spectra of these particles in Au$+$Au collisions, although alternate 
explanations involving mass-dependent partonic energy loss in the hot 
medium are not ruled out.

In summary, we have observed an enhancement of electrons from heavy-flavor 
decays produced in central $d+$Au collisions at \sqsn = 200 GeV.  The 
previously observed suppression of these electrons in central Au$+$Au 
collisions is therefore attributed to hot-nuclear-matter effects.  We find 
that the $\pi^{0}$ and $e^{\pm}_{\rm HF}$ nuclear modification factors 
\raa are consistent within uncertainties in the $\pt$ range where CNM 
effects on both species are small.  In the range where CNM enhancement of 
$e^{\pm}_{\rm HF}$ is significant in $d+$Au, these effects may also be 
apparent in the Au$+$Au data.

%\textbf{*** page break for PRL word count $<$3.5 pages $<$7 columns} 
%\clearpage

%%%%%%%%%%%%%%%%%%%%%%%%%  Acknowledgements 

%\section{Acknowledgements}   % Run-7 long from for PRC, PLB, etc.

We thank the staff of the Collider-Accelerator and Physics
Departments at Brookhaven National Laboratory and the staff of
the other PHENIX participating institutions for their vital
contributions.  We acknowledge support from the 
Office of Nuclear Physics in the
Office of Science of the Department of Energy, the
National Science Foundation, Abilene Christian University
Research Council, Research Foundation of SUNY, and Dean of the
College of Arts and Sciences, Vanderbilt University (U.S.A),
Ministry of Education, Culture, Sports, Science, and Technology
and the Japan Society for the Promotion of Science (Japan),
Conselho Nacional de Desenvolvimento Cient\'{\i}fico e
Tecnol{\'o}gico and Funda\c c{\~a}o de Amparo {\`a} Pesquisa do
Estado de S{\~a}o Paulo (Brazil),
Natural Science Foundation of China (P.~R.~China),
Ministry of Education, Youth and Sports (Czech Republic),
Centre National de la Recherche Scientifique, Commissariat
{\`a} l'{\'E}nergie Atomique, and Institut National de Physique
Nucl{\'e}aire et de Physique des Particules (France),
Bundesministerium f\"ur Bildung und Forschung, Deutscher
Akademischer Austausch Dienst, and Alexander von Humboldt Stiftung (Germany),
Hungarian National Science Fund, OTKA (Hungary), 
Department of Atomic Energy and Department of Science and Technology (India), 
Israel Science Foundation (Israel), 
National Research Foundation and WCU program of the 
Ministry Education Science and Technology (Korea),
Ministry of Education and Science, Russian Academy of Sciences,
Federal Agency of Atomic Energy (Russia),
VR and Wallenberg Foundation (Sweden), 
the U.S. Civilian Research and Development Foundation for the
Independent States of the Former Soviet Union, 
the US-Hungarian Fulbright Foundation for Educational Exchange,
and the US-Israel Binational Science Foundation.

%%%%%%%%%%%%%%%%%%%%%%%%%%%  References 

%\bibliography{ppg131x0}

\begin{thebibliography}{26}
\expandafter\ifx\csname natexlab\endcsname\relax\def\natexlab#1{#1}\fi
\expandafter\ifx\csname bibnamefont\endcsname\relax
  \def\bibnamefont#1{#1}\fi
\expandafter\ifx\csname bibfnamefont\endcsname\relax
  \def\bibfnamefont#1{#1}\fi
\expandafter\ifx\csname citenamefont\endcsname\relax
  \def\citenamefont#1{#1}\fi
\expandafter\ifx\csname url\endcsname\relax
  \def\url#1{\texttt{#1}}\fi
\expandafter\ifx\csname urlprefix\endcsname\relax\def\urlprefix{URL }\fi
\providecommand{\bibinfo}[2]{#2}
\providecommand{\eprint}[2][]{\url{#2}}

\bibitem[{\citenamefont{Adcox et~al.}(2005)}]{PHENIXwhite}
\bibinfo{author}{\bibfnamefont{K.}~\bibnamefont{Adcox}} \bibnamefont{et~al.}
  (\bibinfo{collaboration}{PHENIX Collaboration}), \bibinfo{journal}{Nucl.
  Phys. A} \textbf{\bibinfo{volume}{757}}, \bibinfo{pages}{184}
  (\bibinfo{year}{2005}).

\bibitem[{\citenamefont{Arsene et~al.}(2005)}]{white2}
\bibinfo{author}{\bibfnamefont{I.}~\bibnamefont{Arsene}} \bibnamefont{et~al.}
  (\bibinfo{collaboration}{BRAHMS Collaboration}), \bibinfo{journal}{Nucl.
  Phys. A} \textbf{\bibinfo{volume}{757}}, \bibinfo{pages}{1}
  (\bibinfo{year}{2005}).

\bibitem[{\citenamefont{Back et~al.}(2005)}]{white3}
\bibinfo{author}{\bibfnamefont{B.~B.} \bibnamefont{Back}} \bibnamefont{et~al.}
  (\bibinfo{collaboration}{PHOBOS Collaboration}), \bibinfo{journal}{Nucl.
  Phys. A} \textbf{\bibinfo{volume}{757}}, \bibinfo{pages}{28}
  (\bibinfo{year}{2005}).

\bibitem[{\citenamefont{Adams et~al.}(2005)}]{white4}
\bibinfo{author}{\bibfnamefont{J.}~\bibnamefont{Adams}} \bibnamefont{et~al.}
  (\bibinfo{collaboration}{STAR Collaboration}), \bibinfo{journal}{Nucl. Phys.
  A} \textbf{\bibinfo{volume}{757}}, \bibinfo{pages}{102}
  (\bibinfo{year}{2005}).

\bibitem[{\citenamefont{Adare et~al.}(2010{\natexlab{a}})}]{PPG086}
\bibinfo{author}{\bibfnamefont{A.}~\bibnamefont{Adare}} \bibnamefont{et~al.}
  (\bibinfo{collaboration}{PHENIX Collaboration}), \bibinfo{journal}{Phys. Rev.
  Lett.} \textbf{\bibinfo{volume}{104}}, \bibinfo{pages}{132301}
  (\bibinfo{year}{2010}{\natexlab{a}}).

\bibitem[{\citenamefont{Adare et~al.}(2006)}]{PPG065}
\bibinfo{author}{\bibfnamefont{A.}~\bibnamefont{Adare}} \bibnamefont{et~al.}
  (\bibinfo{collaboration}{PHENIX Collaboration}), \bibinfo{journal}{Phys. Rev.
  Lett.} \textbf{\bibinfo{volume}{97}}, \bibinfo{pages}{252002}
  (\bibinfo{year}{2006}).

\bibitem[{\citenamefont{Adare et~al.}(2007)}]{PPG066}
\bibinfo{author}{\bibfnamefont{A.}~\bibnamefont{Adare}} \bibnamefont{et~al.}
  (\bibinfo{collaboration}{PHENIX Collaboration}), \bibinfo{journal}{Phys. Rev.
  Lett.} \textbf{\bibinfo{volume}{98}}, \bibinfo{pages}{172301}
  (\bibinfo{year}{2007}).

\bibitem[{\citenamefont{Moore and Teaney}(2005)}]{Teaney}
\bibinfo{author}{\bibfnamefont{G.~D.} \bibnamefont{Moore}} \bibnamefont{and}
  \bibinfo{author}{\bibfnamefont{D.}~\bibnamefont{Teaney}},
  \bibinfo{journal}{Phys. Rev. C} \textbf{\bibinfo{volume}{71}},
  \bibinfo{pages}{064904} (\bibinfo{year}{2005}).

\bibitem[{\citenamefont{van Hees et~al.}(2006)\citenamefont{van Hees, Greco,
  and Rapp}}]{vanHees}
\bibinfo{author}{\bibfnamefont{H.}~\bibnamefont{van Hees}},
  \bibinfo{author}{\bibfnamefont{V.}~\bibnamefont{Greco}}, \bibnamefont{and}
  \bibinfo{author}{\bibfnamefont{R.}~\bibnamefont{Rapp}},
  \bibinfo{journal}{Phys. Rev. C} \textbf{\bibinfo{volume}{73}},
  \bibinfo{pages}{034913} (\bibinfo{year}{2006}).

\bibitem[{\citenamefont{Kovtun et~al.}(2005)\citenamefont{Kovtun, Son, and
  Starinets}}]{etaovers}
\bibinfo{author}{\bibfnamefont{P.~K.} \bibnamefont{Kovtun}},
  \bibinfo{author}{\bibfnamefont{D.~T.} \bibnamefont{Son}}, \bibnamefont{and}
  \bibinfo{author}{\bibfnamefont{A.~O.} \bibnamefont{Starinets}},
  \bibinfo{journal}{Phys. Rev. Lett.} \textbf{\bibinfo{volume}{94}},
  \bibinfo{pages}{111601} (\bibinfo{year}{2005}).

\bibitem[{\citenamefont{Vogt}(2003)}]{vogt_kt}
\bibinfo{author}{\bibfnamefont{R.}~\bibnamefont{Vogt}}, \bibinfo{journal}{Int.
  J. Mod. Phys. E} \textbf{\bibinfo{volume}{12}}, \bibinfo{pages}{211}
  (\bibinfo{year}{2003}).

\bibitem[{\citenamefont{Eskola et~al.}(2009)\citenamefont{Eskola, Paukkunen,
  and Salgado}}]{EPS09}
\bibinfo{author}{\bibfnamefont{K.~J.} \bibnamefont{Eskola}},
  \bibinfo{author}{\bibfnamefont{H.}~\bibnamefont{Paukkunen}},
  \bibnamefont{and} \bibinfo{author}{\bibfnamefont{C.~A.}
  \bibnamefont{Salgado}}, \bibinfo{journal}{JHEP}
  \textbf{\bibinfo{volume}{04}}, \bibinfo{pages}{065} (\bibinfo{year}{2009}).

\bibitem[{\citenamefont{Vitev}(2007)}]{CNM_Eloss}
\bibinfo{author}{\bibfnamefont{I.}~\bibnamefont{Vitev}},
  \bibinfo{journal}{Phys. Rev. C} \textbf{\bibinfo{volume}{75}},
  \bibinfo{pages}{064906} (\bibinfo{year}{2007}).

\bibitem[{\citenamefont{Adler et~al.}(2006)}]{PPG030}
\bibinfo{author}{\bibfnamefont{S.~S.} \bibnamefont{Adler}} \bibnamefont{et~al.}
  (\bibinfo{collaboration}{PHENIX Collaboration}), \bibinfo{journal}{Phys. Rev.
  C} \textbf{\bibinfo{volume}{74}}, \bibinfo{pages}{024904}
  (\bibinfo{year}{2006}).

\bibitem[{\citenamefont{Adams et~al.}(2006)}]{STAR_Cronin}
\bibinfo{author}{\bibfnamefont{J.}~\bibnamefont{Adams}} \bibnamefont{et~al.}
  (\bibinfo{collaboration}{STAR Collaboration}), \bibinfo{journal}{Phys. Lett.
  B} \textbf{\bibinfo{volume}{637}}, \bibinfo{pages}{161}
  (\bibinfo{year}{2006}).

\bibitem[{\citenamefont{Adare et~al.}()}]{PPG125}
\bibinfo{author}{\bibfnamefont{A.}~\bibnamefont{Adare}} \bibnamefont{et~al.}
  (\bibinfo{collaboration}{PHENIX Collaboration}),
  \bibinfo{note}{arXiv:1204.0777 (2012) and to be published.}

\bibitem[{\citenamefont{Adare et~al.}(2011{\natexlab{a}})}]{PPG109}
\bibinfo{author}{\bibfnamefont{A.}~\bibnamefont{Adare}} \bibnamefont{et~al.}
  (\bibinfo{collaboration}{PHENIX Collaboration}), \bibinfo{journal}{Phys. Rev.
  Lett.} \textbf{\bibinfo{volume}{107}}, \bibinfo{pages}{142301}
  (\bibinfo{year}{2011}{\natexlab{a}}).

\bibitem[{\citenamefont{Adcox et~al.}(2003)}]{PHENIXNIM}
\bibinfo{author}{\bibfnamefont{K.}~\bibnamefont{Adcox}} \bibnamefont{et~al.}
  (\bibinfo{collaboration}{PHENIX Collaboration}), \bibinfo{journal}{Nucl.
  Instrum. Meth. A} \textbf{\bibinfo{volume}{499}}, \bibinfo{pages}{469}
  (\bibinfo{year}{2003}).

\bibitem[{\citenamefont{Adler et~al.}(2007{\natexlab{a}})}]{PPG044}
\bibinfo{author}{\bibfnamefont{S.~S.} \bibnamefont{Adler}} \bibnamefont{et~al.}
  (\bibinfo{collaboration}{PHENIX Collaboration}), \bibinfo{journal}{Phys. Rev.
  Lett.} \textbf{\bibinfo{volume}{98}}, \bibinfo{pages}{172302}
  (\bibinfo{year}{2007}{\natexlab{a}}).

\bibitem[{\citenamefont{Nakamura et~al.}(2010)}]{PDG}
\bibinfo{author}{\bibfnamefont{K.}~\bibnamefont{Nakamura}} \bibnamefont{et~al.}
  (\bibinfo{collaboration}{Particle Data Group Collaboration}),
  \bibinfo{journal}{Journ. Phys. G} \textbf{\bibinfo{volume}{37}},
  \bibinfo{pages}{075021} (\bibinfo{year}{2010}).

\bibitem[{\citenamefont{Adler et~al.}(2007{\natexlab{b}})}]{PPG060}
\bibinfo{author}{\bibfnamefont{S.~S.} \bibnamefont{Adler}} \bibnamefont{et~al.}
  (\bibinfo{collaboration}{PHENIX Collaboration}), \bibinfo{journal}{Phys. Rev.
  Lett.} \textbf{\bibinfo{volume}{98}}, \bibinfo{pages}{012002}
  (\bibinfo{year}{2007}{\natexlab{b}}).

\bibitem[{\citenamefont{Adare et~al.}(2010{\natexlab{b}})}]{PPG097}
\bibinfo{author}{\bibfnamefont{A.}~\bibnamefont{Adare}} \bibnamefont{et~al.}
  (\bibinfo{collaboration}{PHENIX Collaboration}), \bibinfo{journal}{Phys. Rev.
  D} \textbf{\bibinfo{volume}{82}}, \bibinfo{pages}{012001}
  (\bibinfo{year}{2010}{\natexlab{b}}).

\bibitem[{\citenamefont{Adare et~al.}(2011{\natexlab{b}})}]{PPG077}
\bibinfo{author}{\bibfnamefont{A.}~\bibnamefont{Adare}} \bibnamefont{et~al.}
  (\bibinfo{collaboration}{PHENIX Collaboration}), \bibinfo{journal}{Phys. Rev.
  C} \textbf{\bibinfo{volume}{84}}, \bibinfo{pages}{044905}
  (\bibinfo{year}{2011}{\natexlab{b}}).

\bibitem[{\citenamefont{Cacciari et~al.}(2005)\citenamefont{Cacciari, Nason,
  and Vogt}}]{pp_fit}
\bibinfo{author}{\bibfnamefont{M.}~\bibnamefont{Cacciari}},
  \bibinfo{author}{\bibfnamefont{P.}~\bibnamefont{Nason}}, \bibnamefont{and}
  \bibinfo{author}{\bibfnamefont{R.}~\bibnamefont{Vogt}},
  \bibinfo{journal}{Phys. Rev. Lett.} \textbf{\bibinfo{volume}{95}},
  \bibinfo{pages}{122001} (\bibinfo{year}{2005}).

\bibitem[{\citenamefont{Adare et~al.}(2009)}]{PPG094}
\bibinfo{author}{\bibfnamefont{A.}~\bibnamefont{Adare}} \bibnamefont{et~al.}
  (\bibinfo{collaboration}{PHENIX Collaboration}), \bibinfo{journal}{Phys. Rev.
  Lett.} \textbf{\bibinfo{volume}{103}}, \bibinfo{pages}{082002}
  (\bibinfo{year}{2009}).

\bibitem[{\citenamefont{Adare et~al.}(2008)}]{PPG080}
\bibinfo{author}{\bibfnamefont{A.}~\bibnamefont{Adare}} \bibnamefont{et~al.}
  (\bibinfo{collaboration}{PHENIX Collaboration}), \bibinfo{journal}{Phys. Rev.
  Lett} \textbf{\bibinfo{volume}{101}}, \bibinfo{pages}{232301}
  (\bibinfo{year}{2008}).

\end{thebibliography}

\end{document}